\documentclass{article}
\usepackage{amsfonts}
\usepackage{amsmath}

\input{tcilatex}
\begin{document}

\title{Comment on "Approximate solutions of the Dirac equation for the
Rosen-Morse potential including the spin-orbit centrifugal term"}
\author{A. Ghoumaid, F. Benamira and L. Guechi \\
Laboratoire de Physique Th\'{e}orique, D\'{e}partement de Physique, \and %
Facult\'{e} des Sciences Exactes, Universit\'{e} des Fr\`{e}res Mentouri,
Constantine, \and Route d'Ain El Bey, Constantine, Algeria}
\maketitle

\begin{abstract}
It is shown that the application of the Nikiforov-Uvarov method by Ikhdair
for solving the Dirac equation with the radial Rosen-Morse potential plus
the spin-orbit centrifugal term is inadequate because the required
conditions are not satisfied. The energy spectra given are incorrect \ and
the wave functions are not physically acceptable. We clarify the problem and
prove that the spinor wave functions are expressed in terms of the
generalized hypergeometric functions $_{2}F_{1}(a,b,c;z)$. The energy
eigenvalues for the bound states are given by the solution of a
transcendental equation involving the hypergeometric function.

PACS: 03.65.Fd, 03.65.Ge, 03.65.Db

Keywords: Dirac equation; Spin symmetry; Pseudospin symmetry; Rosen-Morse
potential; Bound states.
\end{abstract}

Recent paper \cite{Ikhdair} paid attention to the bound state problem of the
Dirac equation for the radial Rosen-Morse potential plus the spin-orbit
centrifugal term under the conditions of the spin and pseudospin symmetries
by means of the Nikiforov-Uvarov method (NU) \cite{Nikiforov} . This problem
\ has been reduced to the solution of the respective Schr\"{o}dinger-like
bound state equations%
\begin{eqnarray}
&&\left[ -\frac{d^{2}}{dr^{2}}+\frac{\kappa (\kappa +1)}{r^{2}}+\frac{1}{%
\hbar ^{2}c^{2}}\left( Mc^{2}+E_{n_{r},\kappa }-C_{s}\right) \Sigma (r)%
\right] F_{n_{r},\kappa }(r)  \notag \\
&=&\frac{1}{\hbar ^{2}c^{2}}\left[ E_{n_{r},\kappa
}^{2}-M^{2}c^{4}+C_{s}\left( Mc^{2}-E_{n_{r},\kappa }\right) \right]
F_{n_{r},\kappa }(r);\text{\ \ }F_{n_{r},\kappa }(0)=F_{n_{r},\kappa
}(\infty )=0,  \notag \\
&&  \label{a.1}
\end{eqnarray}%
(see eq.(20)in Ref.\cite{Ikhdair}) and

\begin{eqnarray}
&&\left[ -\frac{d^{2}}{dr^{2}}+\frac{\kappa (\kappa -1)}{r^{2}}-\frac{1}{%
\hbar ^{2}c^{2}}\left( Mc^{2}-E_{n_{r},\kappa }+C_{ps}\right) \Delta (r)%
\right] G_{n_{r},\kappa }(r)  \notag \\
&=&\frac{1}{\hbar ^{2}c^{2}}\left[ E_{n_{r},\kappa
}^{2}-M^{2}c^{4}-C_{ps}\left( Mc^{2}+E_{n_{r},\kappa }\right) \right]
G_{n_{r},\kappa }(r);\text{\ \ }G_{n_{r},\kappa }(0)=G_{n_{r},\kappa
}(\infty )=0,  \notag \\
&&  \label{a.2}
\end{eqnarray}%
(see eq. (22) in Ref.\cite{Ikhdair}) where $M$ is the mass of the particle, $%
E_{n_{r},\kappa }$ is the bound state energy, $\kappa $ denotes the
spin-orbit quantum number, $C_{s}$ and $C_{ps}$ are two constants and the
functions $\Sigma (r)$ and $\Delta (r)$ are identical and represent \ a
certain phenomenological external potential which has been taken in the
Rosen-Morse form

\begin{equation}
\Sigma (r)=\Delta (r)=-4V_{1}\frac{\exp (-2\alpha r)}{\left( 1+\exp
(-2\alpha r)\right) ^{2}}+V_{2}\frac{1-\exp (-2\alpha r)}{1+\exp (-2\alpha r)%
}.  \label{a.3}
\end{equation}

The use of the following approximation

\begin{equation}
\frac{1}{r^{2}}\approx \frac{1}{r_{e}^{2}}\left[ D_{0}-D_{1}\frac{\exp
(-2\alpha r)}{1+\exp (-2\alpha r)}+D2\left( \frac{\exp (-2\alpha r)}{1+\exp
(-2\alpha r)}\right) ^{2}\right] ,  \label{a.4}
\end{equation}%
and the elementary change of variables

\begin{equation}
\text{\ }z=-\exp (-2\alpha r)\text{\ \ }  \label{a.5}
\end{equation}

reduce eqs. (\ref{a.1}) and (\ref{a.2}) to the Gauss's hypergeometric
differential equations%
\begin{equation}
\left[ z(1-z)\frac{d^{2}}{dz^{2}}+(1-z)\frac{d}{dz}-\frac{\beta
_{1}z^{2}-\beta _{2}z+\varepsilon _{n_{r},\kappa }^{2}}{z(1-z)}\right]
F_{n_{r},\kappa }(z)=0;\text{ \ }F_{n_{r},\kappa }(-1)=F_{n_{r},\kappa
}(0)=0,  \label{a.6}
\end{equation}

\begin{equation}
\left[ z(1-z)\frac{d^{2}}{dz^{2}}+(1-z)\frac{d}{dz}-\frac{\overline{\beta }%
_{1}z^{2}-\overline{\beta }_{2}z+\overline{\varepsilon }_{n_{r},\kappa }^{2}%
}{z(1-z)}\right] G_{n_{r},\kappa }(z)=0;\text{\ }G_{n_{r},\kappa
}(-1)=G_{n_{r},\kappa }(0)=0,  \label{a.7}
\end{equation}%
where the quantities $\varepsilon _{n_{r},\kappa },$ $\beta _{1},\beta _{2},%
\overline{\varepsilon }_{n_{r},\kappa },\overline{\beta }_{1}$ and $%
\overline{\beta }_{2}$ are defined by eqs. (28a), (28b), (28c), (45a), (45b)
and (45c) in Ref.\cite{Ikhdair}, respectively. Note that a minus sign is
missing in the boundary conditions (see eqs. (27) and (44) of Ref.\cite%
{Ikhdair}).

At this stage, unfortunately, the author of the Ref. \cite{Ikhdair} employs
the polynomial method (NU) to solve the differential equations (\ref{a.6})
and (\ref{a.7}). To implement this approach, he gave to expressions of the
polynomial $\sigma (z)$ and of the functions $\rho (z)$ and $\phi \left(
z\right) $ the following canonical forms (see eqs. (29), (37) and (38) in
Ref.\cite{Ikhdair}):

\begin{equation}
\sigma (z)=z(1-z),  \label{a.8}
\end{equation}

\begin{equation}
\rho (z)=z^{2\varepsilon _{n_{r},\kappa }}(1-z)^{2\delta +1},  \label{a.9}
\end{equation}%
and%
\begin{equation}
\phi \left( z\right) =z^{\varepsilon _{n_{r},\kappa }}(1-z)^{\delta +1}.
\label{a.10}
\end{equation}%
for the spin symmetry solution and just replace $\varepsilon _{n_{r},\kappa
} $ and $\delta $ by $\overline{\varepsilon }_{n_{r},\kappa }$ and $\delta
_{1} $ for the pseudospin symmetry solution. Then, he claimed to have
obtained the solutions of eqs. (\ref{a.6}) and (\ref{a.7}) as

\begin{eqnarray}
F_{n_{r},\kappa }(z) &=&\phi \left( z\right) y_{n_{r}}(z)  \notag \\
&=&\mathcal{N}_{n_{r},\kappa }z^{\varepsilon _{n_{r},\kappa }}(1-z)^{\delta
+1}P_{n_{r}}^{(2\varepsilon _{n_{r},\kappa },2\delta +1)}(1-2z),
\label{a.11}
\end{eqnarray}%
and

\begin{eqnarray}
G_{n_{r},\kappa }(z) &=&\phi \left( z\right) y_{n_{r}}(z)  \notag \\
&=&\mathcal{N}_{n_{r},\kappa }z^{\overline{\varepsilon }_{n_{r},\kappa
}}(1-z)^{\delta _{1}+1}P_{n_{r}}^{(2\overline{\varepsilon }_{n_{r},\kappa
},2\delta _{1}+1)}(1-2z),  \label{a.12}
\end{eqnarray}%
where $P_{n_{r}}^{(\mu ,\nu )}(x)$ is the Jacobi polynomial.

However, there are some serious points that invalidate these solutions.
First, they do not satisfy the boundary conditions $F_{n_{r},\kappa }(-1)=0$
and $G_{n_{r},\kappa }(-1)=0$ . So they are not physically acceptable. Next,
it should be noted that the weight $\rho (z)$ does not satisfy the condition

\begin{equation}
\left. \sigma (z)\rho (z)z^{k}\right\vert _{a}^{b}=0,\text{ \ \ \ \ \ \ }%
(k=0,1,...),  \label{a.13}
\end{equation}%
(see theorem on the orthogonality of polynomials of hypergeometric-type \ in
Ref. \cite{Nikiforov}, eq. (17), p. 29 ). Here $\left( a,b\right) \equiv
\left( -1,0\right) $. Then, the hypergeometric functions $y_{n_{r}}(z)$ are
not orthogonal polynomials on the interval $\left( -1,0\right) $ and in this
case, one can not extract the eigenvalues of the energy from the equation
(see Ref. \cite{Nikiforov}, eq. (13), p. 9 ), 
\begin{equation}
\lambda _{n_{r}}+n_{r}\tau ^{\prime }+\frac{n_{r}(n_{r}-1)}{2}\sigma
^{\prime \prime }=0;\text{ \ \ \ }(n_{r}=0,1,...).  \label{a.14}
\end{equation}%
Therefore, we must discard the above solutions entirely and the energy
equations given by (34) and (47) in Ref. \cite{Ikhdair} are not correct.

Since the polynomial method (NU) cannot be applied to the discussion of the
radial Rosen-Morse problem, we otherwise proceed to determine the energy
spectrum and the corresponding wave functions. In the case of spin symmetry,
we look for the solution of the equation (\ref{a.6}) in the form%
\begin{equation}
F_{n_{r},\kappa }(z)=z^{\mu }(1-z)^{\delta }y_{n_{r}}(z),  \label{a.15}
\end{equation}%
in which, on account of boundary conditions, $\mu $ has to be positive and $%
\delta $ may be a real quantity. Substituting (\ref{a.15}) into (\ref{a.6})
and taking 
\begin{equation}
\mu =\varepsilon _{n_{r},\kappa },  \label{a.16}
\end{equation}%
and%
\begin{equation}
\delta _{\pm }=\frac{1}{2}\pm \sqrt{\frac{1}{4}+\beta _{1}-\beta
_{2}+\varepsilon _{n_{r},\kappa }^{2}},  \label{a.17}
\end{equation}%
we obtain for $y_{n_{r}}(z)$ the differential equation%
\begin{equation}
\left\{ z(1-z)\frac{d^{2}}{dz^{2}}+\left[ 2\mu +1-\left( 2\mu +2\delta
+1\right) z\right] \frac{d}{dz}-\left( \mu +\delta \right) ^{2}+\beta
_{1}\right\} y_{n_{r}}(z)=0.  \label{a.18}
\end{equation}%
The solution of this equation is the hypergeometric function%
\begin{equation}
y_{n_{r}}(z)=\mathcal{N}\text{ }_{2}F_{1}\left( \mu +\delta _{\pm }+\sqrt{%
\beta _{1}},\mu +\delta _{\pm }-\sqrt{\beta _{1}},2\mu +1,z\right) .
\label{a.19}
\end{equation}%
Now, taking into account the formulas (see Ref. \cite{Gradshtein}, eqs.
(9.131), p. 1043),%
\begin{equation}
_{2}F_{1}(\alpha ,\beta ,\gamma ;z)=(1-z)^{\gamma -\alpha -\beta }\text{ }%
_{2}F_{1}(\gamma -\alpha ,\gamma -\beta ,\gamma ;z),  \label{a.20}
\end{equation}

\begin{equation}
_{2}F_{1}(\alpha ,\beta ,\gamma ;z)=(1-z)^{-\alpha }\text{ }_{2}F_{1}\left(
\alpha ,\gamma -\beta ,\gamma ;\frac{z}{z-1}\right) ,  \label{a.21}
\end{equation}%
the upper component of the radial spinor wave function has the following
expression%
\begin{eqnarray}
F_{n_{r},\kappa }(r) &=&\mathcal{N}\left( -e^{-2\alpha r}\right)
^{\varepsilon _{n_{r},\kappa }}(1+e^{-2\alpha r})^{-\varepsilon
_{n_{r},\kappa }-\sqrt{\beta _{1}}}  \notag \\
&&\times \text{ }_{2}F_{1}\left( \varepsilon _{n_{r},\kappa }+\delta _{+}+%
\sqrt{\beta _{1}},\varepsilon _{n_{r},\kappa }-\delta _{+}+\sqrt{\beta _{1}}%
+1,2\varepsilon _{n_{r},\kappa }+1,\frac{1}{e^{2\alpha r}+1}\right) ,  \notag
\\
&&  \label{a.22}
\end{eqnarray}%
where $\mathcal{N}$ is a constant factor. The solution (\ref{a.22}) fulfills
the boundary condition $F_{n_{r},\kappa }(0)=0$, when%
\begin{equation}
_{2}F_{1}\left( \varepsilon _{n_{r},\kappa }+\delta _{+}+\sqrt{\beta _{1}}%
,\varepsilon _{n_{r},\kappa }-\delta _{+}+\sqrt{\beta _{1}}+1,2\varepsilon
_{n_{r},\kappa }+1,\frac{1}{2}\right) =0.  \label{a.23}
\end{equation}%
Then, the energy values for the bound states are given by the solution of
this transcendental equation (\ref{a.23}) which can be solved numerically.

In order to obtain the pseudosymmetry solution, we proceed similarly as for
the previous case and obtain the lower component of radial spinor wave
function%
\begin{eqnarray}
G_{n_{r},\kappa }(r) &=&\overline{\mathcal{N}}\left( -e^{-2\alpha r}\right)
^{\overline{\varepsilon }_{n_{r},\kappa }}(1+e^{-2\alpha r})^{-\overline{%
\varepsilon }_{n_{r},\kappa }-\sqrt{\overline{\beta }_{1}}}  \notag \\
&&\times \text{ }_{2}F_{1}\left( \overline{\varepsilon }_{n_{r},\kappa }+%
\overline{\delta }_{+}+\sqrt{\overline{\beta }_{1}},\overline{\varepsilon }%
_{n_{r},\kappa }-\overline{\delta }_{+}+\sqrt{\overline{\beta }_{1}}+1,2%
\overline{\varepsilon }_{n_{r},\kappa }+1,\frac{1}{e^{2\alpha r}+1}\right) ,
\notag \\
&&  \label{a.24}
\end{eqnarray}%
where%
\begin{equation}
\overline{\delta }_{+}=\frac{1}{2}+\sqrt{\frac{1}{4}+\overline{\beta }_{1}-%
\overline{\beta }_{2}+\overline{\varepsilon }_{n_{r},\kappa }^{2}},
\label{a.25}
\end{equation}%
\ and $\overline{\mathcal{N}}$ is a constant factor. Then, the energy
spectrum can be also found from a numerical solution of the transcendental
equation

\begin{equation}
\text{ }_{2}F_{1}\left( \overline{\varepsilon }_{n_{r},\kappa }+\overline{%
\delta }_{+}+\sqrt{\overline{\beta }_{1}},\overline{\varepsilon }%
_{n_{r},\kappa }-\overline{\delta }_{+}+\sqrt{\overline{\beta }_{1}}+1,2%
\overline{\varepsilon }_{n_{r},\kappa }+1,\frac{1}{2}\right) =0.
\label{a.26}
\end{equation}

In conclusion, given boundary conditions, the problem of solving the
equations (\ref{a.6}) and (\ref{a.7}) does not belong to the class of
quantum mechanical problems when it comes to search for energy levels and
wave functions of a quantum system moving in a field of forces using
classical orthogonal polynomials as eigenfunctions.


\begin{thebibliography}{9}
\bibitem{Ikhdair} S. M. Ikhdair, J. Math. Phys. 51, 023525 (2010)

\bibitem{Nikiforov} A. F.\ Nikiforov and V. B. Uvarov, \textit{Special
Functions of Mathematical Physics} (Birkh\"{a}user, Bassel, 1988).

\bibitem{Gradshtein} I. S. Gradshtein and I. M. Ryzhik, \textit{Tables of
integrals, series and products} (Academic Press, New York, 1965).
\end{thebibliography}
\end{document}